\documentclass[12pt]{article}
\usepackage{graphicx}
\usepackage{pdfpages}
\usepackage{hyperref}
\usepackage{amssymb}
\usepackage{amsmath}
\numberwithin{equation}{section}
\usepackage{color}
\usepackage[normalem]{ulem}
\usepackage{cite}

\usepackage{pgfplots}
\pgfplotsset{width=10cm,compat=1.9}
\usepackage{tikz}
\usetikzlibrary{decorations}

\pgfdeclaredecoration{complete sines}{initial}
{
    \state{initial}[
        width=+0pt,
        next state=sine,
        persistent precomputation={\pgfmathsetmacro\matchinglength{
            \pgfdecoratedinputsegmentlength / int(\pgfdecoratedinputsegmentlength/\pgfdecorationsegmentlength)}
            \setlength{\pgfdecorationsegmentlength}{\matchinglength pt}
        }] {}
    \state{sine}[width=\pgfdecorationsegmentlength]{
        \pgfpathsine{\pgfpoint{0.25\pgfdecorationsegmentlength}{0.5\pgfdecorationsegmentamplitude}}
        \pgfpathcosine{\pgfpoint{0.25\pgfdecorationsegmentlength}{-0.5\pgfdecorationsegmentamplitude}}
        \pgfpathsine{\pgfpoint{0.25\pgfdecorationsegmentlength}{-0.5\pgfdecorationsegmentamplitude}}
        \pgfpathcosine{\pgfpoint{0.25\pgfdecorationsegmentlength}{0.5\pgfdecorationsegmentamplitude}}
}
    \state{final}{}
}

\usepackage[normalem]{ulem}

\newcommand{\be}{\begin{equation}}
\newcommand{\ee}{\end{equation}}

\begin{document}

\begin{center}
{\Large \bf{Instant Cosmology}}
\\

\vspace{6mm}

\textit{Nissan Itzhaki and Uri Peleg}
\break \break
  School of Physics and Astronomy, Tel Aviv University, Ramat Aviv, 69978, Israel
\end{center}

\vspace{5mm}

\begin{abstract}

Instant Folded Strings (IFSs) are unconventional light strings that emerge when the string coupling increases with time. A particularly intriguing property of IFSs, especially relevant to cosmology, is that they violate the Null Energy Condition (NEC). In this paper, we begin to explore their cosmological effects. We find that NEC violation by IFSs is significantly suppressed in an expanding universe, leading to a universe that resembles our own, comprising matter, radiation, and dark energy. Upon closer examination, these components exhibit subtle, nonstandard traits that could be experimentally tested in the future. Notably, the origin of dark energy stems not only from the potential, as is usually the case, but also from the derivative of the potential with respect to the dilaton. This paves the way for a new approach to realizing inflation within string theory, addressing the Dine-Seiberg problem associated with dilaton stabilization, and perhaps even hinting at a novel mechanism to tackle the cosmological constant problem. Conversely, in a contracting universe, the effects of IFSs are amplified, making bouncing cosmologies a natural and prevalent outcome.

%Instant Folded Strings (IFSs) are nonstandard light strings that appear when the string coupling grows with time. An interesting feature of IFSs, that is particularly relevant for cosmology, is that they violate the Null Energy Condition (NEC). 
%In this paper, we begin to explore their cosmological effects.
%We find that in an expanding universe, their NEC violation is highly suppressed, producing a universe similar to ours, which includes matter, radiation, and dark energy. A closer look reveals that all three components exhibit subtle nonstandard characteristics that might be testable in the future. Most notably, the contribution to dark energy arises not only from the dilaton's potential (as is typically the case) but also from its slope. 
%This opens up new avenues for realizing inflation in string theory, addressing the Dine-Seiberg problem associated with the dilaton stabilization and perhaps even provides a new way to tackle the cosmological constant problem.
%This paves the way for new approaches to realizing inflation within string theory, addressing the Dine-Seiberg problem associated with dilaton stabilization, and perhaps even hinting at a novel mechanism to tackle the cosmological constant problem. In a contracting universe, the IFSs effect is enhanced making a bouncing universe quite common. 

\end{abstract}

\newpage
\section{Introduction}

Instant Folded Strings (IFS) \cite{OG IFS} are nonstandard light strings that emerge when the string coupling increases over time. %These strings can significantly impact cosmological evolution in scenarios where the dilaton, a scalar field related to the string coupling, varies temporally. 
They violate the Null Energy Condition (NEC), making their potential effects on cosmology particularly intriguing. What sets them apart from ordinary light strings is that they cannot be approximated as particle-like, even at large distances. This unique property, which follows from the way they are extended in space and especially time, suggests they could leave distinctive stringy imprints on the cosmos, unlike anything produced by particle physics. This is the motivation behind this paper, in which we take the first steps in exploring the implications of IFSs on Cosmology, which we dub "Instant Cosmology".

%Instant Folded Strings (IFSs) \cite{OG IFS} are nonstandard stringy excitations that make their appearance when the string coupling grows with time. As such they can influence the cosmological evolution in cases in which the dilaton varies with time. IFSs are light and violate the Null Energy Condition (NEC), which makes their potential cosmological effects particularly interesting. At a heuristic level, the most fascinating aspect of IFSs is that, unlike ordinary strings, despite being light they cannot be approximated, not even from afar, by particles. As a result, they may leave unique, stringy imprints on cosmology that cannot be replicated by particles. This is the motivation behind this work in which we start to explore IFSs cosmology, which we dub "Instant Cosmology".

The plan of the paper is as follows.
The next section provides a concise review of IFSs, focusing on their features most relevant to cosmology. In particular, we highlight that in homogeneous cosmological setups, IFSs generate a large negative pressure without contributing to the energy density. In Sec. \ref{EOMs}, we demonstrate that despite being extended and (yet) light,  IFSs can, subject to easily satisfied conditions, be integrated out in a controlled manner. This allows us to derive effective equations of motion for the time-dependent dilaton, $\phi(t)$, and scale factor, $a(t)$. The remainder of the paper explores various properties of these equations. Sec. \ref{Instant slow-roll} demonstrates that the NEC violation induced by IFSs is suppressed in an expanding universe. This fundamental aspect of instant cosmology is not contingent on specific dynamical details; rather, it stems from the Bianchi identity, which manifests as a large friction induced by the IFSs.
%Section 4 shows that the NEC violation induced by IFSs is suppressed by cosmic expansion due to the Bianchi identity that manifests itself through the significant friction generated by the IFSs. 
This friction introduces an attractor mechanism that, for a generic dilaton potential, steers the evolution of $\phi(t)$ and $
a(t)$ toward a slow-roll regime. This simplifies the dynamics considerably and produces a universe consistent with ours, including dark energy, matter, and radiation. 

In Sec. \ref{Late time dark energy}, we examine these components more closely and find that each exhibits nonstandard features potentially testable through observations. A notable finding is that, in instant cosmology, dark energy is sourced not only by the potential, as is typically the case, but also by the derivative of the potential with respect to the dilaton. This leads to a new stringy way for the universe to inflate. We discuss the potential advantages of "instant inflation" over traditional cosmic inflation models. Additionally, we entertain the possibility that IFSs could drive the universe's current acceleration. This scenario provides a novel resolution to the Dine-Seiberg problem \cite{Dine:1985he}, achieving dilaton stabilization through the large friction induced by IFSs, rather than relying on a potential.  Furthermore, we show that the potential’s slope contribution to the effective potential exactly cancels the notorious cosmological constant generated at one loop. Matter fields also behave in nonstandard ways: for instance, their gravitational and inertial masses may differ on cosmological scales, opening new avenues for understanding dark matter.

In a contracting universe, the NEC violation by IFSs is amplified, making bouncing solutions a natural feature of instant cosmology. Sec. \ref{A bouncing universe} deals with some elementary aspects of these bouncing solutions.

\section{A review of IFS}

This section discusses aspects of IFSs that are relevant to cosmology. Most of the section reviews results from \cite{OG IFS,Karinne,Worldsheet Description of IFS,Itzhaki:2021scf,When strings surprise}. 

We start by considering the simplest background in which IFSs make an appearance: a time-like linear dialton with a spatial direction 
\be\label{twod} ds^2 = - dt^2 + dx^2, \quad \phi = Q t. \ee
We consider the case in which $Q$ is positive which means that the singularity, associated with the blow-up of the string coupling, is in the future. We take the dilaton slope to be small, $Q\ll 1$, which suggests that, at least naively, supergravity should be a good approximation at low energies. 

The time-like linear dilaton implies that the central charge associated with this background is smaller than $2$ (or $3$ in the supersymmetric case). There are various ways to embed this in superstring theory. Restricting to compact time-independent manifolds we can, in the spirit of \cite{Gepner:1987qi},  multiply the supersymmetric version of (\ref{twod}) with five ${\cal N}=2$ minimal models which can be described by $SU(2)_{k_i}/U(1)$ coset CFT. Criticality and the fact that the $k_i$'s are integers means that $Q$ cannot be arbitrarily small in the setup. However, it can be small enough, $Q^2\sim 10^{-6}$, to expect low-energy effective action to be a good approximation. 
Relaxing one of the conditions and allowing the extra directions to be non-compact time-independent or compact time-dependent it is easy to see that $Q^2$
can be arbitrarily small. Maybe the simplest way to illustrate this is to replace one of the five ${\cal N}=2$ minimal models with a cigar CFT $SL(2)_k/U(1)$. Since $k$ is continuous $Q$ can be arbitrarily small. 

\begin{figure}
\includegraphics[width=12cm]{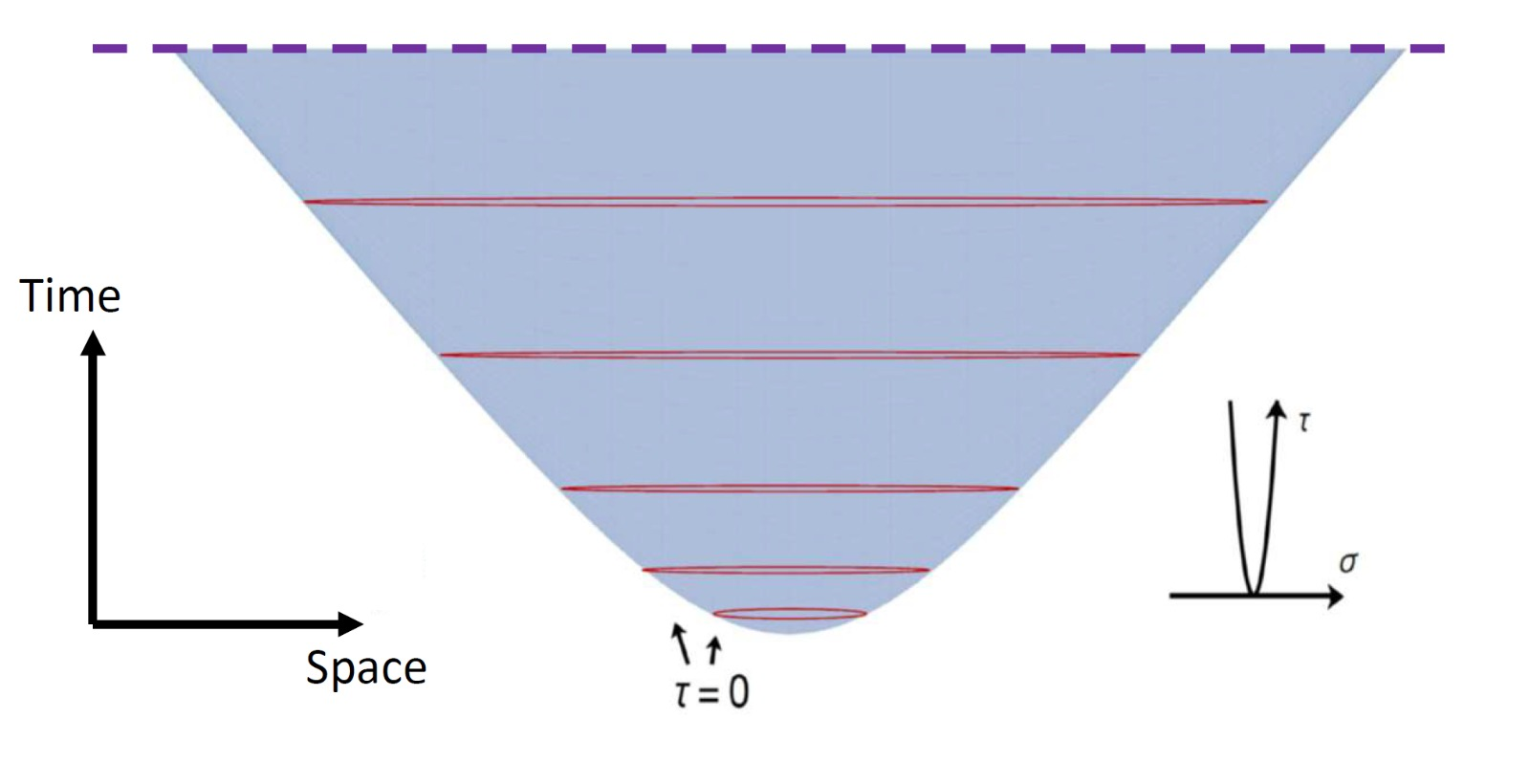}
\centering
\caption{The IFS solution.  A closed folded string is classically created at a specific moment in time and undergoes expansion as it evolves toward a singularity in the future (marked by the dashed purple line).}
\label{fig:IFS}
\end{figure}

At any rate, the surprising aspect of string theory in the background  (\ref{twod}) is that, for any positive $Q$, no matter how small,  supergravity is not a good approximation even at low energies. The reason is that the background (\ref{twod}) includes new light stringy degrees of freedom that can radically modify the low energy physics.  These are the IFSs. Classically they are described by  the following solution to the equation of motion and Virasoro constraints 
\be 
\begin{split}
&t(\sigma,\tau) = t_0 +  Q\ln \left(\frac{1}{2}\cosh \left(\frac{\sigma}{Q}\right) + \frac{1}{2}\cosh\left(\frac{\tau}{Q}\right)\right), \\
&x(\sigma,\tau)=x_0+\sigma,  \label{ifss}
\end{split}
\ee
where we work with $\alpha'=1$ and  $-\infty < \sigma, \tau, < \infty.$
The target space interpretation of the solution is of a closed folded string which is created at $t=t_0$ and $x=x_0$. The size of the folded string expands rapidly, with the fold (corresponding to the $\tau=0$ world sheet slice) following a space-like trajectory that asymptotically approaches a null trajectory (see figure (\ref{fig:IFS})).

An exact CFT description of an IFS was given in  \cite{Worldsheet Description of IFS}
and was used to calculate the IFS production rate\footnote{It is likely that a careful analysis of the one-loop determinant around the classical solution, in the spirit of \cite{Affleck:1981bma} (see also  \cite{Gordon:2014aba,Gordon:2016ldj}), should suffice for the calculation of the production rate.}
\be\label{rate}
\Gamma_{IFS}(t_0)\sim \frac{Q^2}{g_s(t_0)^2}.
\ee
The factor of $Q^2$ implies that, as expected, IFSs are not created when $Q=0$.
The factor of $1/g^2$ is also natural since the IFSs are created classically, implying that the IFSs creation should affect the sphere partition function. Indeed an agreement between (\ref{rate}) and the relevant partition function \cite{Giribet:2011zx} was found in \cite{When strings surprise}. 
The Bekenstein-Hawking entropy associated with near extremal NS5-branes provides yet another, indirect, test of (\ref{rate}) \cite{Worldsheet Description of IFS}.

The observation that an IFS is created classically in an instant, combined with the fact that, as far as fundamental strings are concerned, the background (\ref{twod}) is invariant under time translation, implies that at any time the total energy (and momentum) of an IFS vanishes. This was confirmed through a direct calculation \cite{Karinne}, demonstrating that supergravity is not a valid approximation at low energies as it does not include IFSs that are light and can alter the low-energy dynamics considerably, and in ways that light particles cannot. Put differently, standard light strings look like particles. An IFS, on the other hand, is light, but it does not look like a particle. Even not from afar.  This rather heuristic point is in a sense the reason why, as we demonstrate below, IFSs lead to such non-standard cosmology.

While the total energy of an IFS vanishes its energy-momentum tensor is far from trivial. The bulk of the IFS has positive energy density induced, as usual, by the tension of the folded string. Since the total energy vanishes, this positive bulk energy is compensated by a negative energy at the fold. As the string gets larger the energy at the fold becomes more negative, which means that an IFS violates the NEC. As a result, in backgrounds such as (\ref{twod}), in which IFSs are created homogeneously the cosmological NEC  is violated, $\rho_{IFS}+p_{IFS}<0.$ Since the total energy of a single IFS vanishes the energy density must vanish as well, $\rho_{IFS}=0$, and the NEC violation of an IFS implies that $p_{IFS}$ is negative. Indeed a short calculation gives  \cite{Itzhaki:2021scf} 
\be
p_{IFS}\sim - \Gamma_{IFS} \tau_{IFS}^2 \sim -\frac{ Q^2 }{g_s^2}  \tau_{IFS}^2,
\ee
where $\tau_{IFS}$ is the IFS life-time.

In principle, the exact CFT description of \cite{Worldsheet Description of IFS} can be used to calculate $\tau_{IFS}$. This is a rather complicated calculation involving higher point functions in FZZT-branes \cite{Fateev:2000ik,Teschner:2000md} and is beyond the scope of this paper. Luckily, one can approximate $\tau_{IFS}$ semi-classically by estimating its splitting rate. As the size of the IFS grows linearly with time it is natural  to estimate that 
$
\tau_{IFS}\sim \frac{1}{g_s}
$
which gives
\be
\rho_{IFS}=0,~~~~p_{IFS}=-\frac{\gamma_2}{g_s^4}Q^2,
\ee
where $\gamma_2$ is an order $1$ constant that we cannot determine. 

Similar considerations imply that in four-dimensional time-like linear dilaton
\be ds^2 = - dt^2 + dx^2 +dy^2 +dz^2 +\mbox{compact}, \quad \phi = Q t \label{eq:TLD},\ee 
where the IFS can be stretched in any of the spatial directions, we have
\be\label{four}
\rho_{IFS}=0,~~~~p_{IFS}=-\frac{\gamma_4}{g_s^4}Q^2.
\ee
Again, $\gamma_4$ is a constant of order $1$ which we cannot determine, at least at the moment. Luckily, most of the interesting results below do not depend much on $\gamma_4$. 

The $1/g_s^4$ in (\ref{four})  implies that the backreaction of the IFSs on the time-like linear dilaton background is quite large, and that 
to be able to calculate it we should consider  IFSs  in FRW background with time-dependent dilaton 
\be ds^2 = -dt^2 + a^2(t)\left( \frac{dr^2}{1 - k r^2}+ r^2 d\Omega^2\right), ~~~\phi(t), \label{FLRW metric}
\ee
which is the subject of the next section.

\section{The equations of motion}
\label{EOMs}
%\label{IFS Cosmology}
In this section, we derive the cosmological equation of motions associated with (\ref{FLRW metric}) taking into account the IFSs. In effective field theory terms, we wish to find the effective equation of motions for $a(t)$ and $\phi(t)$ after integrating out the IFSs.  Usually, we can trust such  effective equations of motion when the curvature and the string coupling are small, which roughly speaking means in cosmology that
\be\label{zx}
g_s\ll 1, ~~~~H\ll 1,
\ee
with  $H=\dot{a}/a$.

It turns out that there is an extra condition here. The effective equations of motion for $a(t)$ and $\phi(t)$ can be trusted when, on top of (\ref{zx}), we have
\be\label{po}
g_s \gg H.
\ee
The standard condition (\ref{zx}) is a result of integrating out heavy degrees of freedom, such as massive stringy modes and D-branes. Given that IFSs are light and extended, it is not too surprising that integrating them out in a controlled manner requires an extra, rather unusual, condition.

Let us see how (\ref{po}) comes about.
To find the effective equations of motion for $a(t)$ and $\phi(t)$ in the background (\ref{FLRW metric}) we should be able to generalize (\ref{four}) and find $p_{IFS}$ in these backgrounds. There are, however, two obstacles to achieving this goal. In the background (\ref{FLRW metric}), unlike in the time-like linear dilaton background (\ref{eq:TLD}), we do not know the exact IFS solution that determines its energy-momentum tensor, and we also lack an exact CFT description of IFSs in (\ref{FLRW metric}) that would allow us to calculate the IFS production rate. Nevertheless, we now argue that when the lifetime of the IFS is significantly shorter than the Hubble time, 
\be
\tau_{IFS}  \ll 1/H~~~~\leftrightarrow~~~~g_s\gg H,
\ee 
reliable approximations for both the production rate and the shape of the IFS can be utilized to approximate $p_{IFS}$.

The reason is that, while the IFS is an extended object, its production is a local process that takes time of the order of $Q$, which for $Q\ll1 $, is much smaller than the length scale set by the curvature and/or the second derivative of $\phi$.  $\Gamma_{IFS}$, for example, is determined by the dilaton (and its derivative) at the creation point of the IFS (\ref{rate}). This implies that even though we do not have an exact CFT description of IFSs in the background (\ref{FLRW metric}), it is natural to expect that, as long as $\partial_{\mu}\phi$ is small, time-like  and points to the future, a good approximation is 
\be\label{ki}
\Gamma_{IFS}=\frac{\partial\phi^2}{g_s^2}\Theta(\dot{\phi}),
\ee
where $\Theta(\dot{\phi})$, is the standard theta function which reflects the fact that IFSs are created only when $\dot{\phi}>0.$

Similarly the IFS solution in the background (\ref{FLRW metric}) deviates from (\ref{ifss}). This deviation becomes significant at length scales of the order to the curvature length scale. In maximally symmetric situations, such as \cite{Gubser:2002tv},  one can find the exact string solution even when the string is larger than the curvature scale.  However, in less symmetric situations this is not an easy task, and we were not able to find the exact solutions in (\ref{FLRW metric}). Thus we have to limit ourselves to cases in which the size of the IFS is smaller than $1/H$. 
As discussed in the previous section, at finite string coupling the lifetime and size of an IFS scales like $1/g_s$, which means that for $g_s\gg H$ the IFS solution is well approximated by (\ref{ifss}).  Combining this with (\ref{ki}) we conclude that, subject to (\ref{po})\footnote{In two dimensions, this condition might be easier to relax \cite{Meir:2024unc} by leveraging the cigar puncture solution found in \cite{Brustein:2021qkj}, though its validity is debatable \cite{dk} (for more recent discussion see  \cite{Krishnan:2024zax}). From this perspective (\ref{po}) is a way to evade this debate.}, the pressure induced by the IFSs in the background (\ref{FLRW metric}) is
\be\label{pr} p_{IFS} = -\frac{\gamma}{3 g_s^4}(\partial \phi)^2 \Theta(\dot{\phi}),  \ee
where $\gamma$ is an undetermined constant of order $1$, and the factor of $3$ is included to simplify the cosmological equation of motions presented below.

What about $\rho_{IFS}$? In the time-like linear dilaton background $\rho_{IFS}$ vanishes because of translation invariance in $t$. The background  (\ref{FLRW metric}) does not respect this symmetry, and there is no reason to expect $\rho_{IFS}$ to vanish.  A simple macroscopic way to determine  $\rho_{IFS}$, which does not involve the details of the IFS solution,  is the Bianchi identity, which for fundamental strings in the string frame takes the familiar form
\be \dot{\rho}_{IFS}+3H(\rho_{IFS}+p_{IFS})=0, \label{Bia}\ee
where  $\rho_{IFS}$ and $p_{IFS}$ are the energy density and pressure associated with the IFSs and their decay products. 

Since $p_{IFS}<0$ the Bianchi identity implies that the expansion of the universe tends to increase $\rho_{IFS}$  and by doing so to suppress the NEC violation induced by the IFSs. The cosmological equations of motion, presented below reflect this.  Similarly, a contracting universe tends to amplify the NEC violation caused by the IFSs.
Hence, in a sense, IFSs are particularly well-suited for generating a bouncing cosmology that does not contradict observations in an obvious way. As the universe contracts, IFSs increasingly violate the NEC, eventually triggering an expansion of the universe. During this expansion phase, NEC violation is gradually suppressed, leaving only subtle imprints — some of which are discussed in the following sections — in the expanding universe.

There is a microscopic way to see that in an expanding (contracting) universe the NEC violation of the IFSs is decreased (increased). In the time-like linear dilaton background, the total energy on an IFS vanishes due to a cancellation between the positive contribution from the bulk of the IFS and the negative contribution from the fold. However, the bulk and folds of the string respond differently to the Hubble parameter and this response leads to an imperfect cancellation. The resulting residual energy is linear in 
$H$. This behavior is detailed in Appendix \ref{Bianchi identity}, where it is shown to be consistent with the Bianchi identity (\ref{Bia}).

Equipped with (\ref{pr}) we are now in a position to discuss the IFS cosmological equations of motion. There are still two related issues we need to address. First the IFS energy-momentum, and as a result $\rho_{IFS}$ and $p_{IFS}$, were considered in the string frame. From a cosmological perspective, it is more natural to consider the equation of motion in the Einstein frame. Hence we have to transform to the Einstein frame. A second issue is that an IFS affects not only the metric (via the negative pressure it induces, (\ref{pr})), but also the dilaton.\footnote{Since it is folded, an IFS does not source the $B$ field.} How to take this into account in the case of a single string was explained in \cite{Dabholkar:1995nc,DGHR}. Since we have many IFSs we have to integrate the result of \cite{Dabholkar:1995nc,DGHR} taking into account the IFS shape, its estimated lifetime, and production rate. 
In homogeneous cases, this is fairly straightforward and is done in Appendix \ref{derivation of the cosmological equations}, where we also 
convert from world-sheet conventions to cosmological conventions in which the dilaton is canonically normalized and the potential appears in a standard way. This involve rescaling of the dilaton, $\phi\to \kappa \phi/ \sqrt{2}$ (so that the string coupling is $g_s=e^{\phi}\to e^{\kappa \phi/ \sqrt{2}}$), and potential, $V \to  \kappa^{2}g_s^2V$ and yields the following equations of motion
\begin{subequations}
\begin{gather}
3H^2 = \kappa^2 \rho_{tot}  \equiv \kappa^2 \left(\rho_r + \frac{1}{2}\dot{\phi}^2 + V(\phi)\right)  - \frac{3k}{a^2}, \label{a} \\
\ddot{\phi}+3H\dot{\phi} + V'(\phi) =  -\frac{\kappa\gamma}{\sqrt{2}}   \frac{\dot{\phi}^2}{g_s^{2}} \Theta(\dot{\phi}),\label{b} \\ \dot{\rho}_{r} + 4H\rho_{r} = \gamma \frac{\dot{\phi}^2}{g_s^{2}}\left(H+\frac{\kappa}{\sqrt{2}}\dot{\phi}  \right) \Theta(\dot{\phi}) , \label{c} 
\end{gather}\label{IFSD}
\end{subequations}
where, as usual, $\kappa^2=8\pi G_N$, and we also rescaled $\gamma \to 2\gamma/\kappa^2$ for convenience.
$\rho_{r}$ represents the energy density associated with the radiation the IFSs decay to. 
From a phenomenological perspective, $\rho_r$ can be interpreted as dark radiation.

Subject to the restriction (\ref{po}) (and (\ref{zx})), these equations describe how IFSs modify the dynamics associated with backgrounds of the form (\ref{FLRW metric}). 
In the rest of the paper, we explore some rather basic properties of (\ref{IFSD}). The effect of the IFSs is most transparent in (\ref{b}). 
The left-hand side of this equation is familiar:  the $\ddot{\phi}$ term follows from the canonically normalized kinetic term, $3H\dot{\phi}$ is the Hubble friction term, and we also have the "external force" term due to the slope of the potential, $V'$.  The right-hand side includes the new term induced by the  IFSs. This term is present only when $\dot{\phi}>0$, since only then IFSs are produced.  As  $V'>0$, eventually gives $\dot{\phi}<0$ we focus on $V'<0$. Because the IFSs term 
is negative it increases the friction. As it scales like $1/g_s$ it does so in an extreme way, especially in an expanding universe in which the IFS term enhances dramatically the Hubble friction. Consequently 
when $V'<0$, (\ref{IFSD}) admits, for {\em generic} $V$, a slow-roll approximation, which we discuss next.

\section{Instant slow-roll}
\label{Instant slow-roll}

At weak coupling, $g_s\ll 1$, a generic dilaton potential with $V'<0$ satisfies 
\be\label{con} -V'(\phi) \gg  \kappa g_s^2 |V(\phi)|, ~~~\mbox{and} ~~~ -V'(\phi) \gg\frac{g_s^2}{\kappa}|V''(\phi)|. \ee
As hinted by the discussion above, and shown in detail in Appendix \ref{attractor}, under these conditions an attractor mechanism takes over where $\dot{\phi}$ very quickly becomes suppressed by a factor of $g_s$ relative to $H$ and (\ref{b}) is approximated by\footnote{From now on we do not write down the $\Theta(\dot{\phi})$ term as we consider only cases with $\dot{\phi}>0$.}
\be
V'(\phi) \cong  -\frac{\kappa\gamma}{\sqrt{2}}   \frac{\dot{\phi}^2}{g_s^2}. \label{slow-roll}
\ee
As a result (\ref{a}) and (\ref{c}) are approximated by
\be\label{ooa}
3H^2 \cong \kappa^2 \left(\rho_r+ V\right) , 
\ee
and 
\be
\label{ba}
\dot{\rho}_{r} + 4H\rho_{r}   \cong - \frac{\sqrt{2}}{\kappa} H V',
 \ee
which implies that
\be \rho_r(t) = -\frac{1}{\sqrt{8}\kappa}V' + \frac{1}{a^4(t)}\left(\rho_r(t=0)+\frac{1}{\sqrt{8}k}V'\right) + \mathcal{O}( g_s). \label{total radiation density}\ee
Plugging this into (\ref{ooa}) we get
\be 
3H^2 \cong \left(V -\frac{1}{\sqrt{8}\kappa}V'\right)+\frac{1}{a^4(t)}\left(\rho_r(t=0)+\frac{1}{\sqrt{8}\kappa}V'\right) 
. \ee
Note that the one dimensionless parameter we could not determine in the full instant cosmology equation of motion (\ref{IFSD}), $\gamma$,  does not appear in the slow-roll approximation. $\gamma$ arises due to our inability to calculate $\tau_{IFS}$ precisely, as we were only able to rely on the estimate $\tau\sim 1/g_s$. The fact that $\gamma$ does not appear in the slow-roll approximation seems to imply that, at least in an expanding universe, traces of the IFSs are quickly washed away by the slow-roll attractor. Since we live in an expanding universe this would mean that, despite the fact that IFSs are light and cannot be approximated by particles, there is little hope of making contact with experimental observations. 

Indeed, at first glance, it appears that IFS cosmology is rather standard - it involves radiation and dark energy. A closer look reveals that things are in fact more interesting. Unlike in standard cosmology, $V'$ contributes to both the dark energy and radiation components. Its total contribution 
$$ \frac{1}{\sqrt{8}\kappa}V'\left(\frac{1}{a^4(t)} -1\right), $$
offers a clear illustration, consistent with the discussion in the previous section on the Bianchi identity, emphasizing the sharp contrast between the impact of IFSs in a contracting versus an expanding universe. In an expanding universe, the $-1$ eventually dominates the $1/a^4(t)$, leading to a positive contribution (since $V'$ is negative). Conversely, in a contracting universe, the contribution becomes negative. As previously discussed and further demonstrated in Sec. \ref{A bouncing universe}, this behavior is precisely what enables the universe to bounce.  

It is instructive to also think about the contribution of $V'$ to $V_{eff}$ and $\rho_{eff-rad}$ separately.
Since $V'$ is negative the effective radiation 
\be \rho_{eff-rad} \equiv \frac{1}{a^4(t)}\left(\rho_r(t=0)+\frac{1}{\sqrt{8}\kappa}V'\right),\ee
can be negative, which can be attributed to the fact that IFSs violate the NEC. 

The dark energy component, which controls the expansion rate,  
\be\label{veff} V_{eff} \equiv V -\frac{1}{\sqrt{8}\kappa}V' ,\ee  is larger than the standard dark energy,  $V$ \cite{Peebles:1987ek,Wetterich:1987fm,Caldwell:1997ii}. The extra term, $-\frac{1}{\sqrt{8}\kappa}V'$, is induced by the negative pressure of the IFSs. Since it is positive we can easily have situations in which $V$ is negative, but $V_{eff}$ vanishes or is positive. Some aspects of this are discussed in the next section. 

Eq. (\ref{veff}) implies that the ratio between the pressure and energy density associated with the effective potential is
\be\label{js}
w\equiv\frac{p_{eff}}{\rho_{eff}}=-1+g_s C,~~~~C = -2 \sqrt{-\frac{3V'}{\gamma}} \frac{\sqrt{8}\kappa V'- V''}{\left(\sqrt{8}\kappa V- V'\right)^{3/2}},
\ee
which differs from the standard one in two important aspects. First, at weak coupling, $w$ near $-1$ is a robust result that arises naturally for generic potentials, without the need for fine-tuning. Second, $w$ can be smaller than $-1$. The underlying microscopic cause of this is, once again, the violation of the NEC by the IFSs.

Dark energy plays a crucial role in our universe. It appears in late-time cosmology, during inflation, and perhaps even in the form of early dark energy which was proposed as a solution to the Hubble tension (for a recent review on early dark energy and the Hubble tension see \cite{Kamionkowski:2022pkx}). It is, therefore, encouraging that instant cosmology provides a new way to realize dark energy in string theory at weak coupling. This is particularly exciting because instant cosmology fundamentally depends on strings, making it unattainable within the framework of particle physics. Thus, one can hope that future observations could distinguish instant cosmology acceleration from that obtainable in particle physics. We conclude this section with some rather basic comments about instant cosmology and inflation. In the next section, we discuss instant cosmology and late-time dark energy. Instant cosmology and early dark energy will be discussed elsewhere.

\subsection{ Instant inflation} 
There are three advantages of instant inflation over standard slow-roll models of inflation, especially in string theory.\\  
$\bullet$ Instant inflation does not require tuning the potential for the universe to inflate, while inflation is quite delicate in general and in particular in string theory \cite{Baumann:2007np}.
As long as $g_s\ll 1$ a generic potential generates inflation.  Moreover, there is no graceful exit problem \cite{Albrecht:1982wi} in instant inflation since the slow-roll conditions, (\ref{con}), are not met as the dilaton approaches the minimum of $V$.\\
$\bullet$ Instant inflation does not suffer from the overshoot problem \cite{Brustein:1992nk}, a problem that is quite common in small-field models of inflation which are easier to construct in string theory. String theory offers a potential mechanism for addressing the overshoot problem \cite{Itzhaki:2007nk} and incorporates large field models of inflation\cite{Silverstein:2008sg}. Still, in both cases, it seems that the overshoot problem got traded with the initial condition problem. \\
$\bullet$ Instant inflation seems to evade the initial condition problem as well. Loosely speaking the problem is that even if the potential is such that inflation is possible the question that remains is: why start up in the potential? From energetic considerations, this is not a natural initial condition. This issue has been a subject of debate since the early days of inflation \cite{Goldwirth:1991rj}, and remains a topic of discussion (e.g. \cite{Ijjas:2014nta,Brandenberger:2016uzh,Linde:2017pwt}). In instant inflation, the problem, at least in its basic form, does not appear to arise because it is  $V_{eff}$, and not $V$, that inflates the universe, and due to the $V'$ contribution $V_{eff}$ can be large and positive with small or even negative $V$.

Still, for instant inflation to be a vital alternative to standard inflation we need to know how to calculate fluctuations in instant cosmology properly. These appear to involve a detailed understanding of the IFS decay process, which is currently lacking. There may be a clear imprint that distinguishes between instant inflation and other models of inflation. The discussion in \cite{When strings surprise} seems to imply that all decay channels of an IFS include a null mode with a large and {\em negative } energy
\be
E\sim - \frac{1}{g_s},~~~~P=\pm E.
\ee
Since such modes do not appear in standard inflation it should be interesting to explore their imprints on LSS and CMB. This should be a generalization of \cite{Itzhaki:2008ih,Fialkov:2009xm,Maldacena:2015bha} where the cosmological imprints associated with a massive particle present during inflation were considered.

\section{Late time dark energy} 
\label{Late time dark energy}

A more extreme scenario is that IFSs are also responsible for the universe's current acceleration. That is, the dilaton is still rolling slowly. In this section, we discuss three aspects of this scenario at a rather basic level: dilaton stabilization, (dark)-matter, and the CC problem. We hope that a more detailed understanding of these topics could put us in a position to compare this scenario with experimental results.

\subsection{Dilaton stabilization}

Moduli stabilization in string theory has a long history (for a recent review see \cite{McAllister:2023vgy}). We find it instructive to start with the Dine-Seiberg problem \cite{Dine:1985he}.  The problem pointed out by Dine and Seiberg is that quantum corrections generate a potential for the dilaton that vanishes at weak coupling, where calculations can be performed in a controlled manner. This leads to two options. In the first, the dilaton runs toward zero coupling (fig. 2a). Since $\dot{\phi}<0$ IFSs are not created during this runaway and instant cosmology does not alter the conclusion that the dilaton is pushed towards zero coupling. 

In the second, the dilaton is driven towards strong coupling (fig. 2b) where higher-order corrections in perturbation theory and non-perturbative effects can modify the potential significantly. 
Obtaining a minimum away from the strongly coupled region requires tuning of parameters. The minimum is under better control when $V_{min}<0$, yielding an $AdS$ space.  The uplifting of $AdS$ to $dS$ must involve SUSY breaking and is considered to be the more challenging part in the KKLT construction \cite{Kachru:2003aw} which led to some controversy (see e.g \cite{Sethi:2017phn,Bena:2018fqc}). 

Since $\dot{\phi}>0$
instant cosmology seems to be relevant for this case.  %offer a new possibility. We consider cases in which the quantum potential is such that the dilaton is pushed towards strong coupling. So $V'<0$ and instant cosmology is relevant. 
We saw that for a generic potential $\dot{\phi}\sim g_s H$, which implies that at weak coupling the dilaton is barely changing during a Hubble time, effectively it is stabilized.
It is important to highlight the distinction between Hubble friction and IFS friction in this context. Hubble friction, by definition, is proportional to the Hubble parameter making it effective at slowing down the dilaton in the early universe when $H$ was large. However, its influence diminishes at late times as $H$ decreases. In contrast, IFS friction is independent of $H$
rather it scales like $1/g_s^2$ and so is most effective at weak coupling. Consequently, in such a scenario, it is natural to suspect that we likely live in a weakly coupled region.

\begin{figure}
\includegraphics[width=16cm]{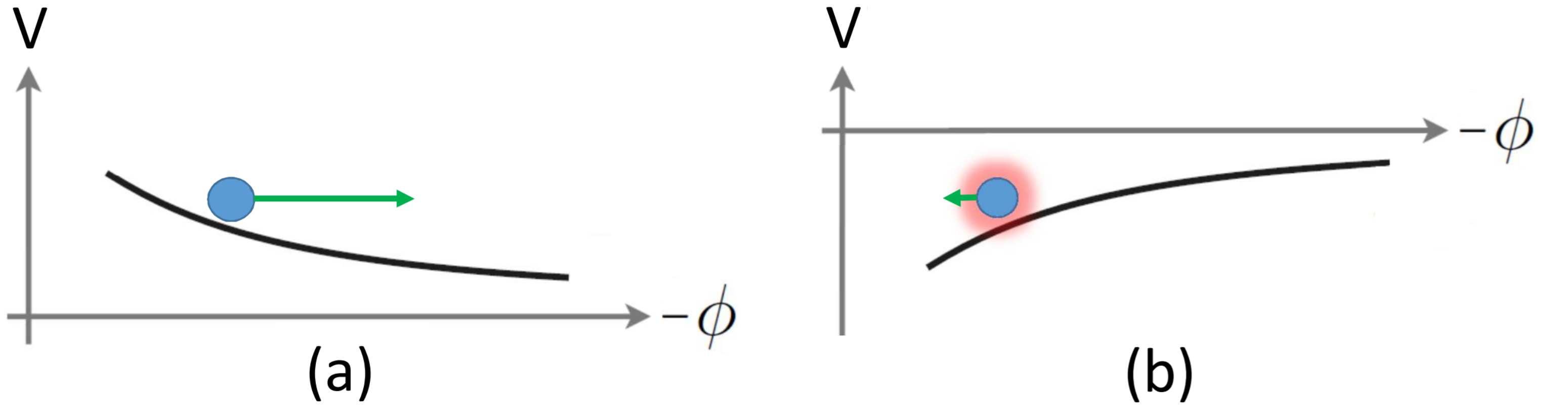}
\centering
\caption{The Dine-Seiberg problem and IFSs. The runaway case (a) is not affected by the IFSs (since $\dot{\phi}<0$). However when the dilaton is driven towards strong coupling (b) it starts to glow as it emits IFSs which slows it down significantly, effectively leading to its stabilization. Since the IFSs friction scales like $1/g_s$ this pseudo stabilization is most effective at weak coupling.  }
\label{fig:potential}
\end{figure}

Another comment worth making is about uplifting. As discussed above, uplifting  $AdS$ to $dS$ is the most challenging aspect of the KKLT construction. Here the uplifting is happening due to the contribution of the slope of the potential, $V'$, to $V_{eff}$. As follows from (\ref{veff}) we can have a positive $V_{eff}$, which yields an approximated dS, even with negative $V$ - the IFSs, which break SUSY, do the uplifting. In particular, we could be heading towards a supersymmetric minimum which is described by an $AdS$ space, but due to the $V'$ contribution, we are experiencing an approximated dS. In this sense, the IFSs play the role of the anti $D$3-branes in the KKLT construction. A crucial difference is that once the dilaton approaches the minimum the slow-roll approximation breaks down, and IFSs are not produced in large numbers, and  $V_{eff} \approx V$ (see Fig. 3).
%and we end up living, in the far future, in an $AdS$ space (see fig. 3). 
An argument why having potentials such that $V$ and $V_{eff}$ have the opposite sign is quite common is given below.

An argument against this dilaton stabilization mechanism is the following. There are, of course, other moduli in string theory that do not produce IFSs when varying with time and so they have to be stabilized in the "old-fashioned way", at the minimum of a potential.   It seems strange that the stabilization mechanism of the dilaton is so different than that of the rest of the moduli fields. Based on black hole considerations it was conjectured, however, in \cite{Itzhaki:2023hwm} that the motion of other moduli fields creates different instant extended objects, such as instant folded D-branes. If correct, this would suggest that other moduli are also stabilized by the friction induced by other, as-yet-undiscovered instant objects. This could potentially reshape the border between the string landscape and the string swampland (for reviews on the landscape and swampland see \cite{Palti:2019pca,Agmon:2022thq}).

\begin{figure}\vspace{-1cm}
\includegraphics[width=12.5cm]{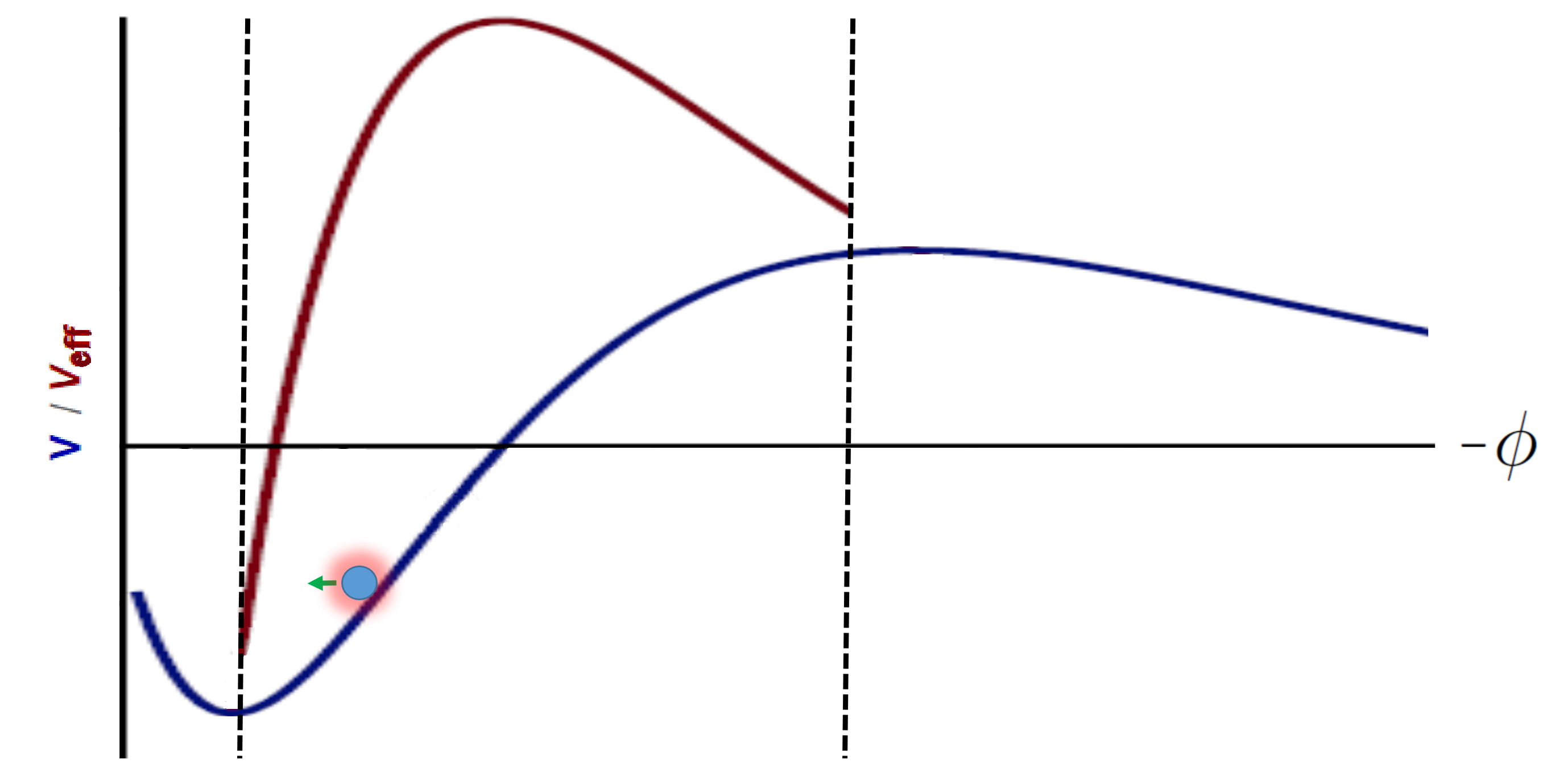}
\vspace{-0.25cm}
\centering
\caption{KKLT and instant cosmology. The blue line represents a typical KKLT potential without the uplifting, so at the minimum there is an $AdS$ space. In the region between the two dashed lines, the slow-roll approximation is valid, and $V_{eff}$, which controls the expansion rate and is represented by the red line, is larger than $V$. We could easily have a situation, represented by the glowing circle, in which $V_{eff}$ is positive while $V$ is negative. In such a case an approximated dS space emerges as we slowly approach the minimum. The negative pressure induced by the IFSs is uplifting $AdS$ to an approximated dS. Eventually near the minimum the dilaton stops glowing and $V_{eff}$ approaches $V$. 
%In this scenario, the universe ultimately evolves into an $AdS$  space in the distant future.
}
\label{fig:kklt}
\end{figure}

\subsection{(Dark)-matter}

Something interesting happens when we add matter to instant cosmology, especially in the slow-roll approximation. The equations of motion are modified in the following way. Eq (\ref{b})  now reads
\be
\ddot{\phi}+3H\dot{\phi}+V'(\phi) = - \frac{\kappa}{\sqrt{2}} \left(\rho_m + \gamma g_s^{-2}\dot{\phi}^2 \right)  
\ee
where $\rho_m$ represents the matter energy density which satisfies 
\be\label{jh}
\rho_m+3H \dot{\rho}_m=\frac{\kappa}{\sqrt{2}} \rho_{m}\dot{\phi}.
\ee
In the slow-roll approximation $H\gg \dot{\phi}$ and so (\ref{jh}) implies that $\rho_m$ acts just like standard matter,
\be\label{sd}
\rho_m(t)=\frac{\rho_m(t=0)}{a(t)^3}. 
\ee
The unusual thing that happens is that $\rho_r$ screens (\ref{sd}). To see this we note that now in the slow-roll limit
\be \dot{\phi} = g_s \sqrt{ -\frac{1}{\gamma} \left(\frac{\sqrt{2}}{\kappa }V'+\rho_m\right)},\ee
which means that 
\be \dot{\rho}_r + 4H\rho_r \cong -H\left(\frac{\sqrt{2}}{\kappa }V'+ \rho_m\right), \label{SR rc}\ee
where, as before, $\rho_r$ represents the energy density associated with the radiation the IFSs decay to. 
Combining this with (\ref{jh}) gives
\be \rho_r(t) \cong - \frac{V'}{\sqrt{8}\kappa} -  \frac{\rho_m(0)}{a^3(t)} + \frac{1}{a^4(t)}\left(\rho_r +  \rho_m + \frac{V'}{\sqrt{8}\kappa} \right)_{t=0},\ee
which implies  that $\rho_m$
modifies $\rho_r(t)$ in two non-trivial ways. First, it changes the dark radiation component that scales like $1/a(t)^4$. This might not be too surprising at this stage since we have already seen that $V'$ does the same.  What is more surprising is that it adds a matter component that scales like $1/a(t)^3$. This extra matter component exactly cancels (\ref{sd}). Thus as far as gravity goes the matter field we added acts like radiation.

Since matter clearly plays a role in the cosmological evolution of our universe, this appears to contradict observations, seemingly ruling out this scenario. However, the discussion above relies on an implicit assumption: that the mass in the string frame is independent of the dilaton. By relaxing this assumption and considering a more general mass term, we arrive at the following equations of motion
\begin{subequations}
\begin{gather}
3H^2 = \kappa^2 \rho_{tot}  \equiv \kappa^2 \left(\rho_r+\rho_m + \frac{1}{2}\dot{\phi}^2 + V(\phi)\right)  - \frac{3k}{a^2}, \label{a^} \\
\ddot{\phi}+3H\dot{\phi} + V'(\phi) = - \frac{\kappa}{\sqrt{2}} (\beta\rho_m + \gamma g_s^{-2}\dot{\phi}^2), \label{b^} \\
\dot{\rho}_{r} + 4H\rho_{r} = \gamma g_s^{-2}\left(H+\frac{\kappa}{\sqrt{2}}\dot{\phi} \right) \dot{\phi}^2, \label{c^} \\
\dot{\rho}_{m} + 3H\rho_{m} = \frac{\kappa}{\sqrt{2}} \beta\rho_{m}\dot{\phi},\label{d^}\end{gather}\label{IFSDG}\end{subequations}
where
\be \beta = \frac{\sqrt{2}}{\kappa}\partial_\phi\ln(m_E(\phi)) = 1 + \frac{\sqrt{2}}{\kappa}\partial_\phi\ln(m_S(\phi)).\ee
Similar considerations to those above give, in  the instant slow-roll approximation
\be \frac{3H^2}{\kappa^2} = \rho_{tot}  \cong \underbrace{V - \frac{\kappa}{\sqrt{8}\kappa}V'}_{V_{eff}}  + \underbrace{\frac{(1-\beta)}{a^3(t)} \rho_m |_{t=0}}_{\rho_{eff-mat}}+ \underbrace{\frac{1}{a^4(t)}\left(\rho_r + \beta \rho_m + \frac{V'}{\sqrt{8}\kappa}\right)_{t=0}}_{\rho_{eff-rad}}  - \frac{3k}{a^2}, \ee
which does include a matter contribution. 

Interestingly, within the framework of homogeneous cosmology, the gravitational effect of matter in this scenario is effectively rescaled by a factor of $(1-\beta) $
\be\label{cra}
G^{matter}_N \to G^{matter}_N (1-\beta).
\ee
$\beta=0$ is the standard matter, where the mass in the Einstein frame is independent of the dilaton. $\beta=1$ is the case discussed at the beginning of this subsection in which the mass in the string frame does not depend on the dilaton and the gravitational effect of the matter is completely screened. 

Another notable case is  $\beta\sim -4$ for which there appears to be no cosmological need (at the homogeneous level) for dark matter as the standard model matter accounts for the necessary gravitational backreaction.  The existence of dark matter is supported, however, by evidence across a range of cosmic length scales - from galaxy rotation curves (10–100 kpc) all the way to the CMB and LSS (Gpc) - each contributing to the overall case for its existence. Therefore, to properly test whether this is a true alternative to dark matter, a deeper understanding of instant cosmology at the inhomogeneous level is required — an understanding that we currently lack.

\subsection{The CC problem}

%Broadly speaking there are two approaches to the CC problem. The first is a CC, or more accurately dark energy, relaxation mechanism.  

%Despite much effort, since the beautiful mechanism proposed long ago by Abbott \cite{Abbott:1984qf}, which unfortunately suffers from the empty universe problem, progress in this direction has been slow. Most of the proposed relaxation mechanisms do not rely on string theory (see however \cite{Feng:2000if}).
%The second approach involves anthropic considerations \cite{Weinberg:1987dv} which appears to fit neatly into string theory  \cite{Bousso:2000xa}.  

We wish to present a short argument that might be viewed as an indication that instant cosmology could provide a new approach to the CC problem.  Consider a typical situation where the CC problem arises in its most basic form.  The potential vanishes or is very small at the classical level and SUSY is broken at some relatively large scale, $\Lambda_{SUSY}$. In such a case a large CC, 
is expected to be generated at one-loop $
C \Lambda_{SUSY}^4
$. Since the torus partition function does not depend on $g_s$, the constant $C$ does not depend on $g_s$ either.  
Hence, in the Einstein frame, with the cosmological conventions we use ($g^{String}_{\mu\nu}=e^{\sqrt{2}\kappa\phi} g^{Einstein}_{\mu\nu}$), the one-loop effective potential  reads
\be
V^{(1)}=C\Lambda_{SUSY}^4 e^{\sqrt{8}\kappa \phi}.
\ee
For negative $C$ (case (b) in fig. 1) we have
$V'<0$, and interestingly enough 
\be
V^{(1)}_{eff} \equiv V -\frac{1}{\sqrt{8}\kappa}V' =0.
\ee
That is, the IFSs contribution to the dark energy, $-\frac{1}{\sqrt{8}\kappa}V'$, automatically cancels the one-loop CC. 

There are corrections to this cancellation. The slow-roll approximation is not exact,  and higher loop corrections do not cancel in this fashion. Since both are of the order of $g_s^2$ we expect to find dark energy that scales like $g_s^2 \Lambda_{SUSY}^4$.  The nice aspect of this is that, at weak coupling, there is a large separation of scales between $\Lambda_{SUSY}^4$ and the dark energy scale. However, the separation is far from sufficient: even if we take $\Lambda_{SUSY}$ to be as low as the Tev scale, to agree with observation we need to have $g_s\sim 10^{-30}$. Such a tiny string coupling is still consistent with (\ref{po}), but is ruled out since it implies that $l_s\sim 10^{30} l_{Planck}$ which is way too large. Still, it is possible that combining with other ideas, such as large extra dimensions 
\cite{Arkani-Hamed:1998jmv,Antoniadis:1998ig} or the RS model \cite{Randall:1999ee,Randall:1999vf}, the scale separation could become sufficiently large to agree with observational data. 

Just like in the discussion in the previous subsection, it is the dilaton coupling and the large friction induced by the IFSs that is suppressing the one-loop CC. The idea that the symmetry behind the resolution of the CC problem is the conformal symmetry is not new (see  \cite{Berman:2002kd,tHooft:2006uhw} and references therein). This cancellation may indicate that IFSs provide a realization of these ideas. Of course, this could simply be a coincidence. 
Even if this is merely a one-loop coincidence it appears to have interesting implications for $V_{eff}$ and moduli stabilization. Consider higher loop corrections. At the n-loop a term that scales like $g_s^{2(n-1)}$ is generated. This means that in the Einstein frame, we get  
\be
V^{(n)}\sim \exp(\sqrt{2}\kappa (n+1) \phi),
\ee
and that, for $n>1$,  $V_{eff}^{(n)}$ and $V^{(n)}$ have opposite signs. This makes the scenario described in fig. \ref{fig:kklt} quite plausible. 

\section{A bouncing universe}
\label{A bouncing universe}

The possibility of finding "bouncing solutions" in cosmology — scenarios where a contracting universe transitions, in a controlled way, into expansion — holds profound implications for our understanding of the universe's origin. Indeed these models have been the subject of extensive study over the years (see e.g. \cite{Veneziano:1991ek,Brustein:1994kw,Kaloper:1995tu,Easther:1995ba,Khoury:2001wf,Khoury:2001bz,Brandenberger:2016vhg,Elitzur:2002rt,Brandenberger:2020tcr}).
These implications could extend beyond conceptual insights, potentially having more practical aspects as well.
A recent example is the observation in \cite{Ijjas:2024oqn} that slow contraction smooths and flattens spacetime. 

The most obvious obstacle to finding such solutions is the fact that for a flat universe to bounce the NEC should be violated,\footnote{For $k=1$ a bouncing universe is possible without violating the NEC. For example, a CC with $p+\rho=0$ yields a bouncing universe - dS in global coordinates.} as follows from 
\be 2\dot{H} = -\kappa^2 (p+\rho) + \frac{2k}{a^2}.\ee
Since IFSs violate the NEC and since, as implied by the Bianchi identity, this violation is increased in a contracting universe, it is not surprising that they lead to bouncing solutions. What is somewhat surprising is that a bouncing solution exists in the slow-roll limit in which $\dot{\phi}\sim g_s$ (since $p_{IFS}\sim -\dot{\phi}^2$).
Below we describe the simplest slow roll bouncing solutions.

The slow-roll limit, as described in Sec. \ref{Instant slow-roll}, is not applicable near the bounce since it relies on $\dot{\phi}\ll H$ and near the bounce $H\to 0$.  In order to describe a bounce in the slow-roll limit, a generalization of the discussion in Sec. \ref{Instant slow-roll} is needed.
Luckily the generalization is rather straightforward as both (\ref{slow-roll}) and (\ref{ooa}) are valid when (\ref{con}) is satisfied even when $|H|\lesssim \dot{\phi}$.
The only equation that requires a new consideration is 
(\ref{ba}), which must now include the previously neglected term proportional to $\dot{\phi}$, and reads
\be \dot{\rho}_{r} + 4H\rho_{r} = \gamma g_s^{-2}\left(H+\frac{\kappa}{\sqrt{2}}\dot{\phi} \right) \dot{\phi}^2 \cong - \frac{\sqrt{2}}{\kappa} H V' - V'\dot{\phi}. \label{decay bianchi}\ee
The solution is 
\be \rho_r(t) \cong -\frac{1}{\sqrt{8}\kappa}V' + \frac{1}{a^4(t)}\left[\rho_r+\frac{1}{\sqrt{8}k}V'\right]_{t=0} - \int^t_0 \left(\frac{a(t')}{a(t)}\right)^4\frac{d}{d\phi}\left[V-\frac{1}{\sqrt{8}k}V'\right]\dot{\phi}(t')dt', \label{phidot radiation density}\ee
which  yields the following expression for the total energy density
\be \frac{3H^2}{\kappa^2} =
V_{eff}+\rho_{eff-rad},
\label{total density}\ee
where 
\be V_{eff} \equiv V(\phi) -\frac{1}{\sqrt{8}\kappa}V'(\phi) - \int_0^t \left(\frac{a(t')}{a
(t)}\right)^4\frac{d}{d\phi}\left[V-\frac{1}{\sqrt{8}k}V'\right]\dot{\phi}(t')dt', \label{slow-roll cc}\ee
and 
\be \rho_{eff-rad} \equiv \frac{\rho_*}{a^4(t)}, \quad ~~ \rho_* \equiv \left[\rho_r+\frac{1}{\sqrt{8}k}V'\right]_{t=0}.\ee

\begin{figure}
\includegraphics[width=13cm]{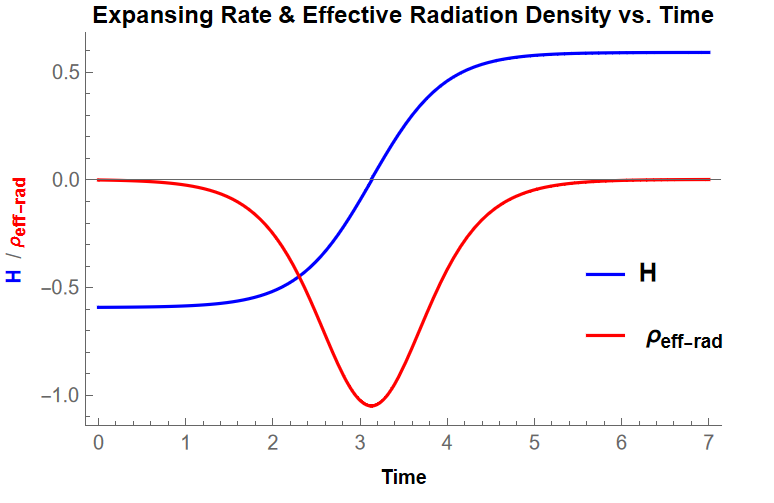}
\centering
\caption{An example for a bouncing solution obtained numerically, initially $H$ is negative, then as the negative energy and pressure build-up, the collapse slows down, and eventually the universe begins to expand, at which point the negative energy density begins to dilute.}
\label{fig:bounce}
\end{figure}
The rate of change of the dark energy
is always subleading relative to the rate of change of $\rho_{eff}$, even when $H\to 0$. The ratio between the two scales like $g_s$.
Hence, for the duration of the bounce we can consider $V_{eff}$ as effectively constant for the duration of the bounce and recast (\ref{total density}) in terms of the effective CC at the time of the bounce, $V_{eff}$, and the initial effective radiation density $\rho_*\equiv \left[\rho_r+\frac{1}{\sqrt{8}k}V'\right]_{t=0}$ where at $t=0$ we set $a=1$.
\be 3\left(\frac{\dot{a}}{a}\right)^2 \cong \kappa^2 \left[V_{eff}+ \frac{\rho_*}{a^4}\right]. \label{simple bounce cosmology}\ee
When $ V_{eff}>0$ and
$ \rho_* <0 $  we have an exact bouncing solution
\be a(t) = \left|\frac{\rho_*}{V_{eff}}\right|^{1/4} \cosh^{1/4}\left[\kappa\sqrt{12 V_{eff}}(t-t_b)\right],\ee
and
\be H(t) \equiv \frac{\dot{a}}{a} = \kappa\sqrt{3V_{eff}} \tanh\left[\kappa\sqrt{12V_{eff}}(t-t_b)\right].\ee
The solution interpolates between a contracting universe, with $H=-\kappa\sqrt{3 V_{eff}}$, at the asymptotic past and expanding universe, with $H=\kappa\sqrt{3 V_{eff}}$, at the asymptotic future. The minimal size of the universe is reached at the time of the bounce $t_b$, where
\be a(t_b) = \left|\frac{\rho_*}{ V_{eff}}\right|^{1/4}_{t=0}. \ee
See fig. \ref{fig:bounce} for a plot of the expansion rate $H$ as well as the total radiation density $\rho_r$ as a function of time, at the vicinity of the bounce.

\section{Conclusions}

This paper aims to initiate the study of the impact of IFSs on cosmology. We considered the homogeneous case and found that
despite being light and extended objects, subject to (\ref{po}), IFSs can be reliably integrated out, yielding effective equations of motion for the time-dependent dilaton and the scale factor. These equations reveal intriguing physics with potential implications both theoretically and phenomenologically.

We hope that future investigations at the inhomogeneous level will deepen our understanding and enable comparisons with experimental data.

\vspace{10mm}

{\bf Acknowledgments:} Much of this work was done while NI was on sabbatical at Princeton. The warm hospitality of the University and the Institute is greatly appreciated. This work was performed in part at Aspen Center for Physics, which is supported by NSF grant PHY-2210452. Work supported in part by the ISF (grant number 256/22). 

\appendix
\section{Derivation of the cosmological equations}
\label{derivation of the cosmological equations}
In this appendix, we derive the instant cosmology equations of motion in general dimension. 
IFS are naturally described in the string frame, using the action
\be S = \frac{1}{2\kappa^2}\int d^D x \sqrt{-g} e^{-2\phi}[R + 2V(\phi) - 4(\nabla \phi)^2] + \text{source},\ee
in $D=d+1$ dimensions. The corresponding EOMs are derived by the variation of the dilaton and metric
\begin{subequations}
\begin{gather}
- 4(\nabla \phi)^2 + 4\nabla^2 \phi + R + 2V(\phi) - V'(\phi)=  0, \label{Dilaton Equation raw} \\
R_{\mu\nu}+2\nabla_\mu\nabla_\nu\phi = \kappa^2 e^{2\phi} (T_{\mu\nu}^{IFS}+T_{\mu\nu}^{SM}) + \frac{1}{2}g_{\mu\nu}V'(\phi) \label{G Equation},\end{gather}\label{1+d DG+IFS}\end{subequations}
where $T_{\mu\nu}^{IFS}$ denotes the energy-momentum tensor of the IFS gas, and $T_{\mu\nu}^{SM}$ is the energy-momentum tensor of radiation and matter.
When written explicitly for the FLRW metric (\ref{FLRW metric}), the dilaton equation (\ref{Dilaton Equation raw}) reads
\be 
 4\dot{\phi}^2 - 4\ddot{\phi} - 4d\left(\frac{\dot{a}}{a}\right)\dot{\phi} + 2d\frac{\ddot{a}}{a}+d(d-1)\left(\frac{\dot{a}}{a}\right)^2 + 2d\frac{k}{a^2} + 2V - V'= 0. \label{Dilaton Equation} \ee
The $R_{00}$ and $g^{ij}R_{ij}$ components of equation (\ref{G Equation}) yield, respectively,
\begin{subequations}
\begin{gather}
-d \frac{\ddot{a}}{a} + 2\ddot{\phi} = \kappa^2 e^{2\phi}(\rho_m + \rho_r) - \frac{1}{2}V' , \label{R_tt Equation}\\
\frac{\ddot{a}}{a} +  (d-1) \left(\frac{\dot{a}}{a}\right)^2  + \frac{2k}{a^2} - 2\left(\frac{\dot{a}}{a}\right)\dot{\phi}= -\frac{\gamma \kappa^2 }{d} e^{-2\phi}\dot{\phi}^2 + \frac{\kappa^2}{d}e^{2\phi}\rho_r + \frac{1}{2}V'.\label{R_ii Equation}
\end{gather}\label{1+d DG+IFS ex}\end{subequations}
Where $\rho_r$ and $\rho_m$ denote the radiation and matter densities, and where we have substituted the IFS pressure into (\ref{R_ii Equation}).

To obtain the analog of FLRW equations in our case of IFS cosmology, we first need to make the transition from the string to the Einstein frame. Since we do not have an effective action description of the IFS gas, a more reliable approach is to apply the conformal transformation from string to Einstein frame at the level of the equations of motion.

The conformal transformation to the Einstein frame involves the change of $a\to a_E =e^{-\frac{2}{d-1}\phi}a$ as well as $dt\to dt_E=e^{-\frac{2}{d-1}\phi} dt$, we deduce that
\be \dot{a}_E \equiv \dot{a} - \frac{2}{d-1}a\dot{\phi} , \quad e^{\frac{2 \phi}{d-1} }\ddot{a}_E \equiv \ddot{a} - \frac{2}{d-1}(\dot{a}\dot{\phi}+a\ddot{\phi}). \label{ES Transformation}\ee
To apply (\ref{ES Transformation}) directly it is convenient to first rearrange our 3 equations (\ref{Dilaton Equation}), (\ref{R_tt Equation}) and (\ref{R_ii Equation}), by taking independent linear combinations, into the form
\begin{subequations}
\begin{gather}
\begin{split}
\frac{d(d-1)}{2}e^{-\frac{4\phi}{d-1}}\left(\frac{\dot{a}_E}{a_E}\right)^2 & \equiv \frac{d(d-1)}{2}\left(\frac{\dot{a}}{a}\right)^2 - 2d\left(\frac{\dot{a}}{a}\right)\dot{\phi} + \frac{2d}{d-1}\dot{\phi}^2  \\  &=\kappa^2 e^{2\phi} (\rho_m + \rho_r) + V + \frac{2}{d-1} \dot{\phi}^2 - d\frac{k}{a^2},
\end{split}
\\
\begin{split}
\frac{d(d-1)}{2} e^{-\frac{4\phi}{d-1}}\frac{\ddot{a}_E}{a_E} & \equiv  \frac{d(d-1)}{2}\frac{\ddot{a}}{a} - 2d\left[\ddot{\phi} + \left(\frac{\dot{a}}{a}\right)\dot{\phi} \right] \\ &= -\frac{\kappa^2}{2}e^{2\phi} [(d-2)\rho_m + (d-1)\rho_r] - 2\dot{\phi}^2 + V + \frac{\gamma \kappa^2}{2} e^{-2\phi}\dot{\phi}^2, 
\end{split}
\\
2\ddot{\phi} + 2d\left(\frac{\dot{a}}{a}\right)\dot{\phi}- 4\dot{\phi}^2 + 2V + \frac{1}{2}(d-1)V'= -\gamma \kappa^2  e^{-2\phi}\dot{\phi}^2 - \kappa^2 e^{2\phi} \rho_m.\end{gather} \label{semi EF}\end{subequations}
The conformal transformation also implies a rescaling of the matter and radiation densities according to the change of the Jacobian in addition, the redefinition $dt\to dt_E=e^{-\frac{2}{d-1}\phi} dt$ implies a corresponding change in the time derivatives of dilaton
\be \rho_E \equiv e^{-2\frac{(d+1)}{(d-1)}\phi}\rho, \quad \dot{\phi}_E \equiv e^{\frac{2\phi}{d-1}} \dot{\phi}, \quad \ddot{\phi}_E \equiv e^{\frac{4\phi}{d-1}} \left(\ddot{\phi}+\frac{2}{d-1}\dot{\phi}^2\right). \label{transformations}\ee
Using (\ref{transformations}) we can rewrite (\ref{semi EF}) in the Einstein frame. For brevity, we will drop the subscript E notation which was previously used to indicate the Einstein frame, from this point onward all of our expressions are written in the Einstein frame
\begin{subequations}
\begin{gather}
\frac{d(d-1)}{2}\left(\frac{\dot{a}}{a}\right)^2 =  \kappa^2(\rho_m + \rho_r) + e^{\frac{4\phi}{d-1}}V + \frac{2}{d-1} \dot{\phi}^2 - 2d\frac{k}{a^2}, \label{3H^2} \\ \frac{d(d-1)}{2}\frac{\ddot{a}}{a} =  -\frac{\kappa^2}{2}[(d-2)\rho_m + (d-1)\rho_r] - 2\dot{\phi}^2 +  e^{\frac{4\phi}{d-1}}V + \frac{\gamma \kappa^2}{2} e^{-2\phi}\dot{\phi}^2, \\
2\ddot{\phi} + 2d\left(\frac{\dot{a}}{a}\right)\dot{\phi}+ \frac{1}{2}(d-1)\left(e^{\frac{4\phi}{d-1}} V \right)' = - \kappa^2 \rho_m -\gamma \kappa^2  e^{-2\phi}\dot{\phi}^2 \label{DE ef}.\end{gather} \label{EF}\end{subequations}
To extract the Bianchi identity from (\ref{EF}), we take the derivative of (\ref{3H^2}) and use the set of equations in (\ref{EF}) to simplify it, ending up with
\be \left[\frac{d}{dt}+(d+1)H\right]\rho_r + \left[\frac{d}{dt}+dH\right]\rho_m = \gamma \kappa^2(H+\dot{\phi})e^{-2\phi}\dot{\phi}^2 + \dot{\phi}\rho_{m}. \label{full bianchi}\ee
In this section, we will focus on the simplest decay scenario where both the bulk and the folds of the IFS quickly decay into radiation (in the form of gravitons). In this case the matter and radiation components of (\ref{full bianchi}) decuple into two separate continuity equations, one for IFS and radiation, and one for matter.
\begin{subequations}
\begin{gather} \dot{\rho}_r +(d+1)H\rho_r = \gamma \kappa^2(H+\dot{\phi})e^{-2\phi}\dot{\phi}^2. \label{IFS-radiation bianchi} \\
 \dot{\rho}_m+dH\rho_m =\dot{\phi}\rho_{m}. \label{matter bianchi} \end{gather} \label{both bianchis}\end{subequations}
When taken together equations (\ref{3H^2}), (\ref{DE ef}) and (\ref{both bianchis}) form a complete set
\begin{subequations}
\begin{gather}
\frac{1}{2} d(d-1) H^2 = \kappa^2 \rho_{tot}  \equiv \kappa^2(\rho_r+\rho_m) + e^{\frac{4d\phi}{d-1}}V + \frac{2}{d-1}\dot{\phi}^2  - d\frac{k}{a^2},  \\
\ddot{\phi} + d H\dot{\phi}+ \frac{1}{4}(d-1)\left(e^{\frac{4\phi}{d-1}} V \right)' = - \frac{1}{2}\kappa^2 \rho_m -\frac{1}{2}\gamma \kappa^2  e^{-2\phi}\dot{\phi}^2,  \\ \dot{\rho}_{r} +  (d+1)H\rho_{r} = \gamma \kappa^2 e^{-2\phi}(H+\dot{\phi}) \dot{\phi}^2,  \\
\dot{\rho}_{m} + d H\rho_{m} = \dot{\phi}\rho_{m}.\end{gather}\label{complete}\end{subequations}
Our equations (\ref{complete}) are written in stringy conventions for which the Einstein frame dilaton-gravity action reads
\be S_E = \frac{1}{2\kappa^2}\int d^D x \sqrt{-g} \left[R + 2e^{\frac{4\phi}{d-1}}V(\phi) - \frac{4}{d-1} (\nabla \phi)^2 \right] + \text{source}.\ee
To match standard cosmological conventions we rescale the dilaton -- such that the kinetic term is canonically normalized, by taking $\phi \to \tilde\phi = \frac{\kappa}{2}\sqrt{d-1}~\phi$. We also rescale the potential by taking, $V \to \tilde V = \kappa^{2}e^{-\frac{4\phi}{d-1}}V$, for it to appear with a standard normalization. 
\be \tilde{S}_E = \int d^D x \sqrt{-g} \left[\frac{1}{2\kappa^2}R - \frac{1}{2}(\nabla \tilde{\phi})^2  + \tilde{V}(\tilde{\phi})\right] + \text{source},\ee
For conveniences, we also rescale, $\gamma \to \tilde{\gamma} = 4\gamma/[\kappa^2(d-1)]$. In the preferred choice of conventions, our cosmological equations (\ref{complete}) now read
\begin{subequations}
\begin{gather}
\frac{1}{2} d(d-1) H^2 = \kappa^2 \rho_{tot}  \equiv \kappa^2 \left[\rho_r+\rho_m + \frac{1}{2}\dot{\phi}^2 + V(\phi)\right]  - d\frac{k}{a^2}, \label{a!} \\
\ddot{\phi}+d H\dot{\phi} +  V'(\phi) = - \frac{\kappa}{\sqrt{d-1}} [\rho_m + \gamma g_s^{-2}\dot{\phi}^2], \label{b!} \\ \dot{\rho}_{r} + (d+1)H\rho_{r} = \gamma \kappa^2 g_s^{-2}\left(H+\frac{\kappa\sqrt{d-1}}{2}\dot{\phi} \right) \dot{\phi}^2, \label{c!} \\
\dot{\rho}_{m} + dH\rho_{m} = \frac{\kappa\sqrt{d-1}}{2} \dot{\phi}\rho_{m}.\label{d!}
\end{gather}\label{final form}
\end{subequations}

\section{An attractor}
\label{attractor}
A straight forward analysis of the dilaton equation (\ref{b!}) at weak coupling $g_s = e^\phi \ll 1$, shows that it exhibits an attractor behavior where $\dot{\phi}$ tends towards a 'slow-roll' value of
\be \dot{\phi}_{SR} \equiv g_s\sqrt{-[V'(\phi)+\rho_m]/(\sqrt{2}\kappa\gamma)}.\ee
when $V' < -\rho_m$.
\iffalse
In which the kinematic terms $\ddot{\phi}$ and $H\dot{\phi}$ in equation (\ref{DE IFS}) become negligible compered to the IFS pressure term $\gamma \kappa^2 e^{-4\phi} \dot{\phi}^2$.
\be 2\ddot{\phi}+6H\dot{\phi} = - V'(\phi) - \rho_m -\gamma \kappa^2 e^{-4\phi} \dot{\phi}^2, \label{DE ef 2}\ee
exhibits an attractor behavior in the weak coupling limit, $g_s = e^\phi \ll 1$, where $\dot{\phi}$ tends towards a 'slow-roll' value of
\be \dot{\phi}_{SR} \equiv \kappa^{-1} e^{2\phi}\sqrt{-[V'(\phi)+\rho_m]/\gamma}.\ee
\fi
The convergence towards this value happens exponentially quickly at a rate of order
\be \gamma_{SR} \equiv \frac{\sqrt{2}\kappa\gamma}{g_s^2} \dot{\phi}_{SR} \cong g_s^{-1} \sqrt{-\sqrt{2}\kappa\gamma [V'(\phi)+\rho_m]} \sim \sqrt{\gamma}g_s^{-1}H.\ee
To see this we can rewrite the dilaton equation (\ref{b!}) in terms of the deviation parameter of $\dot{\phi}$ relative to its slow-roll value,
\be \theta \equiv \dot{\phi}-\dot{\phi}_{SR},\ee
then expand in powers of $g_s$.
\be \dot{\theta} = -\frac{\gamma \kappa}{\sqrt{2} g_s^2} \theta^2 - \frac{1}{g_s} \sqrt{-\sqrt{2}\kappa\gamma [V'(\phi)+\rho_m]} ~ \theta - 3H \theta
+\mathcal{O}(g_s) %\cong -\frac{\gamma}{2 g_s^4} \dot{\phi}_{SR} - 3H. 
\label{critical exp}\ee
It is convenient to recast the equation in terms of the parameters $\gamma_{SR}$ and $\dot{\phi}_{SR}$
\be \dot{\theta}  = - \gamma_{SR}\theta(1+\dot{\phi}_{SR}^{-1} \theta) + \text{subleading}. \label{leading convergance}\ee
Note that during slow-roll we have the following hierarchy of time scales
\be \frac{\dot{\gamma}_{SR}}{\gamma_{SR}},~~\frac{\ddot{\phi}_{SR}}{\dot{\phi}_{SR}} \ll H \ll \gamma_{SR}.\ee
To analyze the rate of convergence to slow-roll, we can view the coefficients $\gamma_{SR}$ and $\dot{\phi}_{SR}$ appearing in (\ref{leading convergance}) as effectively constant relative to the much shorter $t_{SR} \sim \gamma_{SR}^{-1}$ time scale. At this limit the solution to (\ref{leading convergance}) can be approximated by
\be \ln \left(\frac{\theta}{\theta+\dot{\phi}_{SR}}\right) = - \gamma_{SR} t + c. \label{convergance-sol}\ee
Indeed, $\theta = 0$, is an attractor and all initial configurations with $\theta > - \dot{\phi}_{SR}$, meaning $\dot{\phi}>0$, flow to it. For large value $\theta \gg \dot{\phi}_{SR}$ the decay is much more rapid.
Note that (\ref{convergance-sol}) implies that even an arbitrarily large initial value of $\theta$, will still be reduced to a fraction of $\dot{\phi}_{SR}$ within the same typical time frame of just a few $t_{SR} \sim \gamma^{-1}_{SR}$. 

The rapid convergence of any dilaton slop $\dot{\phi} > 0$ to the slow-roll value of $\dot{\phi}_{SR}$ within a time frame of order several $t_{SR} \sim \gamma^{-1}_{SR}$, is a general property of (\ref{leading convergance}).

\iffalse
This is a general property of the ODE, but can be easily seen in the case of constant $\gamma_{SR}$ and $\dot{\phi}_{SR}$ where the exact solution is
\be \ln \left(\frac{\theta}{\theta+\dot{\phi}_{SR}}\right) = - \gamma_{SR} t + c.\ee
Since, $H\gamma_{SR}^{-1} \sim \kappa^{-1} g_s$, the short time it takes for the dilaton to subtle on the slow-roll value is insignificant relative to cosmological time scales. In the case of $g_s \ll 1$ we can accurately model cosmology in the slow-roll limit.
\fi

\section{Bianchi identity at the microscopic level}
\label{Bianchi identity}
Being a fundamental string, an IFS must obey the Bianchi identity in the string frame. In a humongous universe described by the FLRW metric, the Bianchi identity takes the form
\be \dot{\rho}+3H(\rho+p)=0. \label{Bianchi}\ee
A consequence of the Bianchi identity is that the energy density of an IFS gas, and the individual IFSs composing it, is non-vanishing and dependent $H$.
To quantify the correction at the level of a single IFS, we consider a gas of IFSs expanding and diluting as
$\rho=a^{-3}E_{IFS},~ p=a^{-3}PV_{IFS}$, prior to their decay.
For the individual string the Bianchi identity in (\ref{Bianchi}) yields
\be \dot{E}_{IFS}+3H~PV_{IFS}=0. \label{IFS EC}\ee
With $E_{IFS}$ denoting the energy of single IFS and $PV_{IFS}$ its pressure-volume.
The pressure volume associated with a single IFS, smeared isotopically, is given by\footnote{Recall that we are working in $\alpha'=1$ units} 
\be PV_{IFS} = \frac{1}{3}\int dx' T_{11}(t,x') \cong -\frac{2 \Delta t}{3\pi} + \mathcal{O}(H \Delta t^2), \label{IFSPV}\ee
Where $\Delta t \equiv t-t_0$ is the elapsed time since the formation of the IFS.
By substituting (\ref{IFSPV}) into (\ref{IFS EC}), and solving it given the initial condition $E_{IFS}=0$ (at the time of formation), we obtain the leading order correction to the IFS energy%(\ref{EC}).
\be E_{IFS} \cong \frac{H \Delta t^2}{\pi} + \mathcal{O}(H^2 \Delta t^3). \label{EC}\ee
which is completely fixed by the Bianchi identity.

In section (\ref{ansatz}) we show how this correction to the IFS energy arises from the microscopic description of the bulk and folds and their dynamics in an expanding universe, and use it to derive (\ref{IFS EC}) and (\ref{EC}) independently.

In section (\ref{rindler}) we verify this expression in the case of the flat time-like linear dilaton background parametrized in Rindler coordinates, where the exact IFS solution is known. Because the leading order correction is universal and dependent solely on $H$, this serves as an additional independent way to arrive at (\ref{EC}).

\subsection{The bulk and folds during Hubble expansion}
\label{ansatz}
In section \ref{Bianchi identity} above, we have discussed the implications of Bianchi identity at the level of a single IFS. We have seen that the Bianchi identity implies a small correction to the energy of an IFS, proportional to its lifetime time Hubble constant.
The purpose of this section the microscopic origin of this correction and to describe the effect of the Hubble expansion on the bulk and folds. 

Our analysis will be performed in the limit where $Q \ll \Delta t$, such that a non-decaying IFS effectively covers the full future wedge. In this limit, we obtain a precise expression for the energy and pressure derived from the bulk and folds of the IFS.

First, let us begin by considering the bulk of the IFS. The bulk is static, thus its contribution to the energy is twice the string tension, $2T=\frac{1}{2\pi}$, times its proper length. The IFS's proper length is simply the diameter of the light cone, yielding
\be E_{bulk}(t) = \frac{L_{IFS}(t)}{2\pi} = \frac{a(t)}{\pi}\int_{t_0}^t \frac{dt}{a(t)}. \label{IFS bulk energy}\ee
For a short time interval $\Delta t \equiv t-t_0$ relative to cosmological time scales, such that $\dot{H}\Delta t^2 \ll H\Delta t \ll 1$, the leading order correction to the bulk energy can be obtained by considering a constant rate of Hubble expansion $H$
\be E_{bulk}(t) \cong -\frac{1}{2\pi}\left[e^{H(t-t_0)}-1\right] \cong \frac{\Delta t}{\pi} + \frac{H \Delta t^2}{2\pi} +\mathcal{O}\left(\Delta t^3\right) . \label{IFS bulk leading}\ee
The folds of an IFS follow null trajectories and, much like ultra-relativistic particles, they contribute only to the null component of the energy-momentum tensor
\be E_{fold}=-|p_{fold}|. \ee
Much like ultra-relativistic particles, we expect the folds to also experience the Hubble friction's damping effect, decreasing their energy as the universe expands.

More precisely, in the local frame characterized by the clock, $d\tau=a(t)dt$, the fold would feel only the constant force of twice the string tension. 
In this frame, due to the work performed by the tension, the fold's energy will change according to
\be \frac{d}{d\tau}\bar{E}_{fold}(\tau) = -\frac{1}{2\pi}, \ee
The energy conjugate to the self-time $\tau$ is related to energy conjugate to the time $t$ via
$\bar{E}_{fold}(\tau) \equiv a(t) E_{fold}(t)$.
The transition back to the energy conjugate to $t$,
will introduce $H$ precisely in the form of Hubble friction
\be \frac{d}{dt}E_{fold} + H E_{fold} = \frac{d}{d\tau}\bar{E}_{fold}(\tau)= -\frac{1}{2\pi} \label{PTE}.\ee
We can solve (\ref{PTE}) to obtain
\be E_{fold}(t) = -\frac{a^{-1}(t)}{2\pi} \int_{t_0}^t a(t) dt. \label{IFS fold energy}\ee
For a short time interval $\Delta t \equiv t-t_0$ relative to cosmological time scales, such that $\dot{H}\Delta t^2 \ll H\Delta t \ll 1$, the leading order correction to the bulk energy can be obtained by considering a constant rate of Hubble expansion $H$
\be E_{fold}(t) \cong -\frac{1}{2\pi}\left[1-e^{-H(t-t_0)}\right] \cong -\frac{\Delta t}{2\pi} + \frac{H \Delta t^2}{4\pi} +\mathcal{O}\left(\Delta t^3\right) . \label{IFS fold leading}\ee
The total energy of the IFS is the sum of the bulk and fold energies
\be 
\begin{split}
E_{IFS}(t) &= E_{bulk}(t) + 2E_{fold}(t) = \frac{1}{\pi} \left[a(t)\int_{t_0}^t \frac{dt}{a(t)} ~ - ~ a^{-1}(t)\int_{t_0}^t a(t) dt \right] \\ &\cong \frac{H \Delta t^2}{\pi} +\mathcal{O}\left(\Delta t^3\right), 
\end{split}
\label{explicit energy}\ee
which to leading order in the $\dot{H}\Delta t^2 \ll H\Delta t \ll 1$ expansion is precisely the correction in 
The total contribution to the pressure volume, obtaining a negative contribution from both the bulk and folds, is
\be
\begin{split}
PV_{IFS}(t) &= \frac{1}{3}\left[2E_{fold}(t)-E_{bulk}(t)\right] = \frac{1}{3\pi} \left[a(t)\int_{t_0}^t \frac{dt}{a(t)} ~ + ~ a^{-1}(t)\int_{t_0}^t a(t) dt \right] 
\\ &\cong \frac{2\Delta t}{3\pi} +\mathcal{O}\left(\Delta t^3\right).
\end{split}\label{explicit pv}
\ee
Note that equations (\ref{explicit energy}) and (\ref{explicit pv}) in their exact form %combined 
obey the condition of equation (\ref{IFS EC}) precisely -- which is the manifestation of the Bianchi identity at the level of an individual IFS, before its decay.

\subsection{IFS in Rindler coordinates}
\label{rindler}

The simplest nontrivial case in which we can test the bulk and fold energy corrections obtained in section \ref{ansatz} (equations (\ref{explicit pv}) and (\ref{explicit energy}) respectively) is the case of the flat time like linear dilaton expressed in Rindler coordinates.

This simple case is all we need to uniquely determine the lading order correction to the energy, for any FLRW background.
For a short-lived IFS relative to cosmological time scales, such that $\dot{H}\Delta t^2 \ll H\Delta t \ll 1$,
the leading order correction to the energy in (\ref{EC}) is universal and dependent solely on $H$. In particular, it cannot depend on the curvature $R$, since it is non-leading, appearing only at second-order
\be \Delta E_{curvature} \sim R \Delta t^3 =\Delta t^3 (\dot{H}+H^2) \ll \frac{H \Delta t^2}{\pi } \cong E_{IFS}.\ee
This means that if we are only interested in the leading order correction, we can focus on a local neighborhood where space-time is sufficiently flat and ignore curvature corrections. We can preserve $H$ by locally mapping our metric to a patch of flat Rindler space, or equivalently, parametrization of our LIF in Rindler coordinates with the appropriate $H$. 
\iffalse
In flat space, we can write
\be T_{00} = \frac{1}{2\pi\alpha'}\Theta(x_0^2-x_1^2) - \frac{x_0^2+x_1^2}{2\pi\alpha'}\delta(x_0^2-x_1^2),\ee
\be T_{01} = \frac{2x_0x_1}{\pi\alpha'}\delta(x_0^2-x_1^2), \ee
\be T_{11} = -\frac{1}{2\pi\alpha'}\Theta(x_0^2-x_1^2) - \frac{x_0^2+x_1^2}{2\pi\alpha'}\delta(x_0^2-x_1^2).\ee
\fi

We can start with the known energy-momentum tensor in Minkowski coordinates
\be
T_{uv} = \frac{1}{2\pi}\Theta(u)\Theta(v),
\ee
\be  T_{uu} = -\frac{1}{2\pi}\Theta(v)v \delta(u), \quad T_{vv} = -\frac{1}{2\pi}\Theta(u)u  \delta(v),\ee
then make the appropriate coordinate transformation to Rindler coordinates
\be ds^2  = dudv = -dx_0^2+dx_1^2 = -dt^2 + t^2d\phi^2.\ee
In terms of the Rindler coordinates, $x_0 \equiv t\cosh(\phi)-t_0, ~ x_1 \equiv t\sinh(\phi)$, shifted here by the creation time $t_0$ of the IFS. We can then calculate $T_{tt}$ in the Rindler coordinates
\iffalse
\be 
\begin{split}
T_{tt} = &\left(\frac{dx_0}{dt}\right)^2 T_{00} + \left(\frac{dx_1}{dt}\right)^2 T_{11} + 2\frac{dx_0}{dt}\frac{dx_1}{dt}T_{01} = \\ 
&\frac{1}{2\pi\alpha'} \Theta\left(\frac{t^2+t^2_0}{2t_0t} - \cosh(\phi)\right)
- \frac{1}{\pi\alpha'}\delta\left(\frac{t^2+t^2_0}{2t_0t} - \cosh(\phi) \right) ~ \times \\
& \times \frac{e^{-2|\phi|}(t\cosh(\phi) -t_0)\sinh(\phi)}{2t_0} 
%\left[\frac{\cosh(2\phi)(t_0^2+2t_0t\cosh(\phi)+t^2\cosh(2\phi)+8\cosh(\phi)\sinh^2(\phi)(t_0t+t^2\cosh(\phi))}{2t_0t}\right].
\end{split}
\ee
\fi
\be 
\begin{split}
T_{tt} = &\left(\frac{du}{dt}\right)^2 T_{uu} + \left(\frac{dv}{dt}\right)^2 T_{vv} + 2\frac{du}{dt}\frac{dv}{dt}T_{uv} = \\ 
&\frac{1}{2\pi} \Theta\left(\frac{t^2+t^2_0}{2t_0t} - \cosh(\phi)\right)
- \frac{1}{\pi}\delta\left(\frac{t^2+t^2_0}{2t_0t} - \cosh(\phi) \right) ~ \times \\
& \times \frac{e^{-2|\phi|}(t\cosh(\phi) -t_0)\sinh(\phi)}{2t_0} 
%\left[\frac{\cosh(2\phi)(t_0^2+2t_0t\cosh(\phi)+t^2\cosh(2\phi)+8\cosh(\phi)\sinh^2(\phi)(t_0t+t^2\cosh(\phi))}{2t_0t}\right].
\end{split}
\ee
The $\delta\left(\frac{t^2+t^2_0}{2t_0t} - \cosh(\phi) \right)$ term will give us a contribution at the argument's two zero points corresponding to the two folds at
$$\phi_{\pm}= \pm \cosh^{-1}\left(\frac{t^2+t^2_0}{2t_0t}\right)  = \pm \ln \left(\frac{t}{t_0}\right),$$ 
while the $\Theta\left(\frac{t^2+t^2_0}{2t_0t} - \cosh(\phi) \right)$ term will give us the contribution from the bulk.
By integrating $T_{tt}$ over a constant $t$ slice, we find that the total energy contribution of the IFS bulk is
\be 
\begin{split}
E_{bulk}(t) \equiv &\lim_{\epsilon \to 0}\int_{\phi_-+\epsilon}^{\phi_+-\epsilon} d\phi ~ t ~T_{tt} = \frac{t(\phi_+-\phi_-)}{2\pi} = \frac{t}{\pi}\cosh^{-1}\left(\frac{t^2+t^2_0}{2t_0t}\right) 
= \frac{t}{\pi}\ln \left(\frac{t}{t_0}\right),
\end{split} \label{rindler bulk}
\ee
while each fold contributes energy of
\be 
\begin{split}
E_{fold} (t) \equiv &\lim_{\epsilon \to 0}\int_{\phi_\pm-\epsilon}^{\phi_\pm+\epsilon} d\phi ~ t ~ T_{tt} = 
-\frac{e^{-2|\phi_\pm|}}{\pi}\frac{ (t\cosh (\phi_\pm) -t_0)}{2t_0} = \frac{1}{2\pi}\frac{t^2-t_0^2}{2t}.
\end{split} \label{rindler fold}
\ee
Note that our expressions for the bulk and fold energies (\ref{rindler bulk}) and (\ref{rindler fold}), agree precisely with the exact expressions (\ref{IFS bulk energy}) and (\ref{IFS fold energy}) obtained in section \ref{ansatz}, when we substitute $H\equiv t^{-1}$ for the Rindler metric.

By taking $\Delta t = |t-t_0| \ll t_0 $, such that $H=1/t$ can be considered approximately constant, we can determine the leading order correction to the bulk and fold energies. Here we find that
\be E_{bulk}(t) \cong \frac{\Delta t}{\pi}+ \frac{H \Delta t^2}{2\pi} +\mathcal{O}\left(\Delta t^3\right),\ee
\be  E_{fold}(t) = - \frac{\Delta t}{2\pi} + \frac{H \Delta t^2}{4\pi}.\ee
Which in total yields exactly the result of (\ref{EC}),
\be E_{total} = E_{bulk}+ 2E_{fold} \cong \frac{H}{\pi}\Delta t^2 + \mathcal{O}\left(\Delta t^3\right).\ee

\end{document}